\documentclass[aps,prl,reprint,superscriptaddress,twocolumn,nofootinbib]{revtex4-2}

\usepackage{graphicx}
\usepackage{amsmath,amsfonts,amssymb,amsthm}
\usepackage{color}
\usepackage{soul}
\usepackage{physics}
\usepackage{mathrsfs}
\usepackage{mathtools}
\usepackage{physics}

\begin{document}

\title{SWAP and Transpose by displacements, Stabilizer Renyi entropies for continuous variables and qudits and other applications}

\author{Israel Klich}
\affiliation{Department of Physics, University of Virginia, Charlottesville, VA, USA}  

\begin{abstract}
In this note, I highlight a useful formula for the SWAP operator as an average of anti-correlated Heisenberg-Weyl displacements, valid for arbitrary-dimensional quantum systems. As an application I show how the relation can be used to quickly prove normalization identities for the Weyl function, and apply the result to Weyl magic and Wigner magic as the generalization of the recently suggested Renyi Stabilizer entropy to q-dits and CV.
\end{abstract}

\maketitle



The SWAP operation is a fundamental operation in quantum mechanics, that is useful in a variety of contexts. For example, as a quantum gate in quantum computing applications, it is used to facilitate the ability to increase the effective qubit to qubit connectivity needed for error correction. It has been demonstrated experimentally in a variety of systems such as quantum dots \cite{kandel2019coherent} and optical systems \cite{starek2018experimental}. Beyond it's use in quantum computing, is an important tool, for example in the context of quantum entanglement measures, SWAP is naturally used for computing quantities such as the second Renyi entanglement entropy \cite{hastings2010measuring}. 

In this note I discuss a simple formula for the SWAP operator applicable both to qudits as well as continuous variable systems. I  will follow by showing several example of how it can be used. These include Renyi entanglement entropies, and to define appropriate state dependent probability measures on the natural group of displacements (the Heisenberg Weyl group), which can be used to define stabilizer entropy on qudits systems.

Let us start with the SWAP by displacements on arbitrary finite Hilbert space. The SWAP is defined by its action on product states: $SWAP \ket{f}\otimes\ket{g}=\ket{g}\otimes \ket{f}$. To proceed we recall the definition of the phase and shift operators, and the corresponding displacement operators  \cite{vourdas2004quantum}.  
Consider a $d$ dimensional state space, and an orthonormal basis $\ket{j}$, $j=0,..,d-1$. The shift and phase operators are defined as
\begin{eqnarray}&
    X|j\rangle =|j+1\text{ mod } d\rangle \\ & Z|j\rangle =e^{\frac{2\pi  i
   j}{d}}|j\rangle \text{    }.
\end{eqnarray}
For $d=2$, $X$ and $Z$ are simply the appropriate Pauli operators.
For $d>2$ the Heisenberg-Weyl displacement operators are defined as 
 \begin{eqnarray}&
T_{a, b}=e^{-\frac{\pi  i
   a b}{d}}Z^{a}X^{b}\text{  },
\end{eqnarray}
and obey the Heisenberg-Weyl algebra:
 \begin{eqnarray}&
T_{a, b}T_{a', b'}=e^{\frac{i \pi
   (a b'-a' b)}{d}}T_{a + a', b+b'}
\end{eqnarray}
and
 \begin{eqnarray}&
T_{a, b}^{\dag}=T_{-a,- b}.
\end{eqnarray}
The displacement operators are unitary, and may be viewed as a generalization of the Pauli matrices.

{\it SWAP by displacements}:\\ 
The first SWAP by displacements formula I would like to point out is the following:
 \begin{eqnarray}&\label{eq: qudit SWAP}
\frac{1}{d}\Sigma _{a ,b=0}^{d-1}T_{a ,
   b}^{\dagger }\otimes T_{a , b}=SWAP
\end{eqnarray}
acting on $\mathbb{C}^d\otimes \mathbb{C}^d$.

{\it Derivation of \eqref{eq: qudit SWAP}}. It is sufficient to check the relation on basis states:
\begin{eqnarray}& 
    \frac{1}{d}\Sigma _{a ,b}T_{a ,
   b}^{\dagger }\otimes T_{a ,
   b}|j_1,j_2\rangle
   = \\ \nonumber & \frac{1}{d}\Sigma
   _{a ,b}e^{-\frac{2\pi  i
   a \left(j_1-j_2-b\right)}{d}}|j_1
   -b\text{ mod } d,j_2+b\text{ mod }
   d\rangle
\end{eqnarray}
carrying out the $a $ sum, we see that the sum is non vanishing only if $b=j_1-j_2 \text{ mod } d$. It follows that: 
 \begin{eqnarray}&
\frac{1}{d}\Sigma _{a ,b=0}^{d-1}T_{a_1,
   b}^{\dagger }\otimes T_{a , b}|j_1,j_2\rangle=|j_2,j_1\rangle=  \\ \nonumber &  SWAP|j_1,j_2\rangle
\end{eqnarray}
thus we expressed the SWAP operator as a sum of displacements. 

The above result immediately generalizes to multiple qudit systems. Consider a system of $n$ qudits, each has a Hilbert space of dimension $d_i$. Given a vector $\mu
=\left(a_1,b_1;a_2,b_2,\text{...}\text{..}a_n,b_n\right)$, where $(a_i,b_i)\in \mathbb{Z}_{d_i}^{2}$ we can write a generalized displacement as:
\begin{eqnarray}\label{displacement qudit general}
    T_{\mu }=T_{a_1, b_1}\otimes T_{a_2,
   b_2}\otimes \text{...}\otimes
   T_{a_n, b_n}.
\end{eqnarray}
acting on $H=
  \mathbb{C}^{d_1} \otimes...\otimes \mathbb{C}^{d_n}$. Then 
  \begin{eqnarray}&
\frac{1}{ dim(H)}\Sigma_{\mu} T_{\mu}^{\dagger }\otimes T_{\mu}=SWAP
\end{eqnarray}
  acting on $H\otimes H$.

{\it Transpose by displacements}.
A corollary to the SWAP by displacement is the following expression for "Transpose by displacements":
Given a matrix $\rho$ acting on $H$ we have 
\begin{eqnarray}\label{eq:transpose by displacements}
   \frac{1}{dim(H)}\Sigma _{\mu}T_{\mu} \rho T_{\mu}^*=\rho^T,
\end{eqnarray}
where $T_{\mu}^*$ is the complex conjugate of $T_{\mu}$.
The expression follows directly from the SWAP by displacement formula: Vectorize the matrix $\rho$, i.e. $\rho=\sum_{i,j} \rho_{ij}\ket{i}\bra{j} \rightarrow \sum_{i,j} \rho_{ij}\ket{i}\ket{j}$, then apply the "SWAP by displacement" formula and undo the vectorization.

A SWAP by displacement holds also in the continuous variable setting, and can be  viewed as an infinite dimensional limit of \eqref{eq: qudit SWAP}.  Consider a continuous variable mode.
The displacement operator is now defined as: 
\begin{eqnarray}
    D(z)=e^{ z a^{\dagger
   }-z^*a}=e^{-\frac{| z| ^2}{2}}e^{ z
   a^{\dagger }}e^{ -z^*a}
\end{eqnarray}
where $a,a^{\dagger}$ are Boson creation annihilation,  acting on the Fock space spanned by the states $\ket{n}={1\over \sqrt{n!}}a^{\dagger
   n}\ket{0}$, $n=0,1,...$, with $[a,a^{\dagger}]=I$. 
   
The SWAP by displacements now takes the form:
   \begin{eqnarray}\label{CV SWAP identity}
       \int \frac{d^2z}{\pi }D(-z)\otimes
   D(z)=\text{SWAP}
   \end{eqnarray}
where SWAP is a gate that acts on a two-mode Fock space, with the action on product states: 
\begin{eqnarray}
    SWAP\ket{f}_A\ket{g}_B=\ket{g}_A\ket{f}_B.
\end{eqnarray}
acting on $L^2(\mathbb{R})\otimes L^2(\mathbb{R})$.

{\it Derivation of \eqref{CV SWAP identity}}. 
A straightforward calculation allows us to compute explicitly the matrix element of the left hand side of \eqref{CV SWAP identity} between coherent states:
\begin{eqnarray}&\label{CV calc}
    \langle u_1|_A\langle u_2|_B\int
   \frac{d^2z}{\pi }D(-z)\otimes
   D(z)|w_1\rangle_A |w_2\rangle_B= \\ \nonumber & e^{\frac{-1}{2}\left(\left| u_1\right|
   {}^2+\left| u_2\right| {}^2+\left|
   w_1\right| {}^2+\left| w_2\right|
{}^2\right)}e^{u_2^*w_1}e^{u_1^*w_2}=\\ \nonumber & \langle u_1,w_2\rangle\langle u_2,w_1\rangle= \langle u_1|_A\langle u_2|_B SWAP|w_1\rangle_A |w_2\rangle_B.
   \end{eqnarray}
We used that, acting on a coherent state $|w\rangle\equiv D(w)|0\rangle$,  
\begin{eqnarray}
    D(z)|w\rangle =e^{i
   \Im\left(z w^*\right)}|z+w\rangle.
\end{eqnarray}
and that the inner product of coherent states is $\langle z',z\rangle =e^{-\frac{1}{2}|
   z'| ^2}e^{-\frac{1}{2}| z|
   ^2}e^{\overline{z'} z}$. 

As before the above result immediately implies that for several CV modes, we will have the multi-mode SWAP represented as:
\begin{eqnarray}\label{eq:multi qmode SWAP}
    \int \frac{\Pi _id^2z_i}{\pi
   ^N}D\left(-z_1\right)\text{...}D\left(-z_N\right)\otimes   D\left(z_1\right)\text{...}D\left(z_N\right)
\end{eqnarray}

The SWAP by displacement formulas can be used in a variety of contexts. Below I present several examples.

{\it Renyi entanglement entropy.} Entanglement entropy is an important conceptual tool useful for understanding quantum many body systems as well as an important resource for quantum information processing. 
Quantum entanglement is studied across several research fields. For example, in quantum information theory, quantum entanglement serves as a crucial resource for performing computational tasks that are unattainable within classical information theory.   
In condensed matter physics, entanglement is used to diagnose quantum critical phenomena and to analyze the dynamics of strongly correlated quantum systems \cite{eisert2010colloquium}. It is also a key tool for characterizing topological quantum phases of matter that cannot be distinguished by their symmetries \cite{kitaev2006topological,hamma2005ground,levin2006detecting}. 
In quantum field theory, entanglement has a role as a probe for the quantum nature of space-time, black holes and the confinement/ deconfinement phase transitions \cite{nishioka2018entanglement}.

Let us recall the definition of the second Renyi entropy. Consider a state $\rho$ on a Hilbert space $H_A\otimes H_B$. The Renyi entropy associated with subsystem $A$ is defined through the reduced density matrix $\rho_A=Tr_B \rho$, with  the second Renyi entropy being, 
\begin{eqnarray}
    S_2=-\log \tr_A {\rho_A}^2.
\end{eqnarray}
In particular, when $\rho=\ket{\psi}\bra{\psi}$ is a pure state, $S_2$ vanishes if and only if $\ket{\psi}$ is a product state $\ket{\psi}=\ket{\psi_A}\otimes\ket{\psi_B}$ and therefore $S_2$ is a measure of the entanglement in the system.
We now show how $S_2$ on a pure state can be written as an averaging over the Weyl-Heisenberg group. 
It is straightforward to check that the second Renyi entropy can be expressed by using two copies of the wave function $\ket{\psi}\otimes \ket{\psi}$ on the doubled space $(H_A\otimes H_B)\otimes (H_A\otimes H_B)$, with:
\begin{eqnarray}
     \tr_A \rho_A^2=\bra{\psi}\otimes \bra{\psi} SWAP_A \ket{\psi}\otimes \ket{\psi}
\end{eqnarray}
where $SWAP_A$ acts between the two copies of $H_A$. The above formula has been used in many situations, e.g. for accessing Renyi entanglement entropy in quantum Monte Carlo simulations \cite{hastings2010measuring}. Armed with the SWAP-by-displacements expression we immediately have a new expression of $S_2$ using a single copy expectation values of displacements. Using the translations on $H_A$ as defined in Eq. \eqref{displacement qudit general}, we write:
\begin{eqnarray}\label{Renyi entropy}
   S_2=-log  \frac{\Sigma_{\mu} \left| \bra{\psi} T_{\mu}\otimes I_B\ket{\psi}\right|^2}{ dim(H_A)}
\end{eqnarray}
where $I_B$ is the identity on $H_B$ (alternatively,  $I_B$ is viewed as the trivial "zeroth" translation on that space). 

Note that if we take $\ket{\psi}$ to be a product state, $\ket{\psi_A}\otimes\ket{\psi_B}$  in \eqref{Renyi entropy}, $S_2$ should vanish. This immediately suggests that the squared expectation values of the Heisenberg-Weyl translations $T_\mu$ can be associated with a probability distribution. In our next application we will use this property.

{\it  Negativity.}

Let us consier the transpose by displacements eq. \eqref{eq:transpose by displacements}. The expression looks similar to a quantum channel, where $\rho\rightarrow \sum_{\mu} A_{\mu} \rho A_{\mu}^{\dag}$, however in general $T_{\mu}^*\neq T_{\mu}^\dag$, and indeed \eqref{eq:transpose by displacements} is not a quantum channel. It is well known that $\rho\rightarrow \rho^T$ cannot be a proper quantum channel. A proper quantum channel, or Krauss map, has the property that if you apply it to a subsystem of a bigger system, positivity is preserved, but this is not the case for transpose. Indeed partial transpose of the density matrix on non-separable states can have negative eigenvalues and thus reveal entanglement. This observation is the basis for the introduction of quantum negativity as a measure of non-seperability of density matrices. In particular, we have immediately the following formula for a partial transpose:
\begin{eqnarray}
    \rho ^{\text{PT}}=\frac{1}{\dim \left(H_B\right)}\Sigma _{\mu
   _B}I\otimes T_{\mu _B}\rho _ I\otimes T_{\mu _B}^*.
\end{eqnarray}

{\it Application: Weyl magic: a Stabilizer Renyi entropy measure for CV and for q-dits. }

We now show how the above formulae can be used to create entropy measures that are invariant under multimode translations, and hence useful to quantify distance between a given pure state and stabilizer states. 



Stabilizer states are a notable set of states that play a central role in quantum error correction and other applications \cite{gottesman1998heisenberg}. The stabilizer states are derived from qubit computational basis eigenstates using Clifford gates, which are unitary operations transforming any product of Pauli operators into another such product. However, non-Clifford operations are required for universal quantum computation and quantum advantage. This limitation can be overcome by utilizing special states as resources, although this generally incurs higher costs \cite{bravyi2005universal}. Therefore, it is crucial to characterize the "magic" of a state, which refers to the extent to which it deviates from being a stabilizer state. This characterization is important for understanding the cost of preparing the state on a quantum computer and evaluating its utility as a resource for achieving quantum advantage.

These considerations prompted the development of various metrics for quantifying magic. Many of these metrics rely on measuring the distance between a state and the set of stabilizer states and their mixtures. Unfortunately, this set is super-exponentially large, making these metrics impractical for many-body states, except in special cases where rough estimates can be obtained \cite{liu2022many,heimendahl2022axiomatic,ellison2021symmetry,fliss2021knots}. The "mana" metric \cite{veitch2014resource} utilizes the negativity of the discrete Wigner transform of a state. However, as a measure of magic, it is only applicable to systems of qudits with odd prime local Hilbert space dimensions, excluding the commonly used spin-1/2 qubits. Even for qudits, calculating the mana involves operations that scale with the size of the system's Hilbert space, limiting its applicability to relatively small subsystems \cite{sarkar2020characterization,white2021conformal}. Recently, the concept of Rényi entropy of magic for pure states was introduced by Leone, Oliviero and Hamma \cite{leone2022stabilizer}. This approach shows promise \cite{oliviero2022measuring,haug2023quantifying}, although it requires a summation over the exponentially large Pauli group.

Let us recall the stabilizer set up explicitly. The Pauli group $\mathcal{P}_n$ is defined as the set of Pauli strings, together with global $\pm 1,\pm i$ phases. We define the Clifford group $C_n$ as the group of matrices stabilizing the Pauli group. Namely
\begin{eqnarray}
    U\!\in\!  C_n\Leftrightarrow 
   \forall P \!\in\!  P_n\text{  },\text{ 
   }U P U^{\dagger }\!=\!e^{i \phi
   (U,P)}P'(U)\text{     },\text{  }P'\!\in\! P_n
\end{eqnarray}
where $\phi (U,P)$ is a phase. The set of pure stabilizer states is defined as: 
\begin{eqnarray}
S=\left\{U |0\rangle  :\text{  }U\in C_n\right\}.
\end{eqnarray}
Since Clifford operations are known to be non universal, we have the important task of quantifying, given a non stabilizer state resource, the degree of its "non-stabilizernes". 

The proposal of Leone Oliviero and Hamma \cite{leone2022stabilizer} is an elegant measure for non-stabilizarness in the context of the Pauli group. It is based on the observation that 
\begin{eqnarray}
    \frac{1}{ 2^n}\Sigma _{p\in
   \mathcal{P}_n}| \langle \psi |P|\psi
   \rangle | ^2=1.
\end{eqnarray}
It immediately follows that 
\begin{eqnarray}
   \Xi _P(|\psi \rangle )\equiv \frac{ |
   \langle \psi |P|\psi \rangle |
   ^2}{2^n}
\end{eqnarray}
is a probability measure on the n-qubit Pauli group $P_n$. One can now proceed and define associated entropies $S(\Xi(|\psi \rangle)$, which would have the natural property that $S(\Xi(|\psi \rangle)=S(\Xi(U |\psi \rangle)$ for any element of the Clifford group $U\in C_n$.

Let us now turn to defining a corresponding measure for qudits and CV systems.  
The Clifford group $C_n^{d}$ for the n-qudit system is the normalizer of the Weyl-Heisenberg gourp, inside the group of all unitary n-qudits operations. It is defined by the condition:
\begin{eqnarray}
    U\in  C_n^d\Leftrightarrow  \forall \mu
   \text{  }\exists \mu ',\phi \text{  }
   s.t.\text{     }U T_{\mu }U^{\dagger
   }=e^{i \phi }T_{\mu '}\text{     }
\end{eqnarray}
where $\mu,\mu'\in \mathbb{Z}_d^{2n}$ parametrize displacements and $\phi$ is a phase.
In direct analogy with \cite{leone2022stabilizer}, given a pure state $\psi$, let us then define:
\begin{eqnarray}
    p_{\psi }(\mu )\equiv \frac{ \left|
   \langle \psi |T_{\mu }|\psi \rangle
   \right| {}^2}{d^n}
\end{eqnarray}
We now use our SWAP formulas above to show that $p_{\psi }(\mu )$ is a probability measure, induced by $\psi$, on the set of displacements.
Our main task is to prove that for a normalized wavefunction, 
\begin{eqnarray}
    \Sigma _{\mu } p_{\psi }(\mu )=1.
\end{eqnarray}
While this property readily follows from Eq. \eqref{Renyi entropy} by taking $H_A$ to be the entire Hilbert space (and $H_B$ trivial), it is instructive to carry out the calculation explicitly.
The main idea is to write $\left|
   \langle \psi |T_{\mu }|\psi \rangle
   \right| {}^2$ in a doubled space. Thus:
\begin{eqnarray}&
    \Sigma_{\mu\in
   \mathbb{Z}_d^{2n} } p_{\psi }(\mu )=\frac{1}{d^n}\Sigma _{\mu }\langle \psi
   |T_{\mu }^{\dagger }|\psi \rangle
   \langle \psi |T_{\mu }|\psi \rangle
   =\\ \nonumber & \frac{1}{d^n}\Sigma _{\mu
   }\text{Tr}_{H_{d,n}} \left(T_{\mu
   }^{\dagger }|\psi \rangle \langle
   \psi |\right)\text{Tr}_{H_{d,n}}
   \left(T_{\mu }|\psi \rangle \langle
   \psi |\right)=\\ \nonumber & \frac{1}{d^n}\Sigma
   _{\mu }\text{Tr}_{H_{d,n}\otimes
   H_{d,n}} \left((T_{\mu }^{\dagger
   }\otimes T_{\mu })|\psi \rangle \langle \psi |\otimes
   |\psi \rangle \langle \psi
   |\right).
\end{eqnarray}
Now inserting the sum over $\mu$ and using our SWAP relation, we find that
\begin{eqnarray}&
 \nonumber \Sigma_{\mu\in
   \mathbb{Z}_d^{2n} } p_{\psi }(\mu )=   \text{Tr}_{H_{d,n}\otimes H_{d,n}}
   (\text{SWAP} |\psi \rangle \langle
   \psi |\otimes |\psi \rangle \langle
   \psi |)=\\  & \text{Tr}_{H_{d,n}\otimes
   H_{d,n}} ( |\psi \rangle \langle
   \psi |\otimes |\psi \rangle \langle
   \psi |)=1
\end{eqnarray}
as promised. 

Note that in a similar manner, one can show that: 
\begin{eqnarray}
    \Sigma _{\mu } \frac{ \left|
   \langle \psi |T_{\mu }|\phi \rangle
   \right| {}^2}{d^n}=\left|
   \langle \psi |\phi \rangle
   \right| {}^2.
\end{eqnarray}

Given the probability distribution $p_{\psi}$ we can now define a stabilizer Renyi entropy as was done in \cite{wang2023stabilizer}. Defining the Renyi entropy as
\begin{eqnarray}
M_{\alpha }(|\psi \rangle
   )=\frac{1}{1-\alpha }\log   \Sigma_{\mu\in
   \mathbb{Z}_d^{2n} } p_{\psi }(\mu ){}^{\alpha }-\log \text{ 
   }d^n \label{Renyi qudits}
\end{eqnarray}
one can check that it enjoys the desirable properties:\\
(1) For any $U\in C_n^d$, 
$M_{\alpha }(U |\psi \rangle
   )=M_{\alpha }(|\psi \rangle
   )$
   \\ (2) $M_{\alpha }(\psi)=0$ if and only if $\psi\in STAB$.
   \\ (3) additivity $M_{\alpha }(\psi \otimes \phi)=M_{\alpha }(\psi )+M_{\alpha }(\phi)$.
   Thus, indeed $M_{\alpha }$ represents a measure of the non-stabilizerness of a pure state in direct generalization of the result \cite{leone2022stabilizer}.

A detailed analysis of the properties of this measure are presented in \cite{wang2023stabilizer}.

We now turn to the case of continuous variables. 
Consider a system of $N$ q-modes. For CV systems used to encode a Gottesman-Kitaev-Preskil a resource theory has been developed in  \cite{haug2023quantifying} in terms of Wigner negativity, defined as $log \int d^n q d^n p |W_{\sigma}(q,p)|$, where for a quantum state $\sigma$ the Wigner function of is \begin{eqnarray}
    W_{\sigma}=\frac{1}{(2\pi)^n}\int d^n x e^{i p x} \bra{q+\frac{x}{2}}\sigma \ket{q-\frac{x}{2}}.
\end{eqnarray} 
While the system is a CV system, the states are restricted to be GKP states and as such only a small subset of translations (those corresponding to the effective $X$ and $Z$ qubit operations) are required.
Below we show how the Weyl entropy function offers another approach for constructing displacement-invariant measures that can be used for more general CV encodings. 
Let us define the function 
\begin{eqnarray}
    p_{\psi}(z_1,\text{..},z_N)\equiv
   \frac{1}{\pi^N }\left| \tilde{W}_{\psi}\right| {}^2
\end{eqnarray}
where 
\begin{eqnarray}
   \tilde{W}_{\psi}(z_1,\text{..},z_N)\equiv
  \langle \psi
   |D\left(z_1\right)\text{...}D\left(z
   _N\right)|\psi \rangle 
\end{eqnarray}
is called a multimode Weyl function.
The Wigner function can be obtained from a the Weyl function (also known as the characteristic function) by a Fourier transform; see e.g. \cite{vourdas2004quantum} for more details. We note that negativity of the Weyl function does not hold the same place as for the Wigner function, where Hudson's theorem \cite{hudson1974wigner,soto1983wigner,gross2006hudson} and its generalizations state that the Wigner function of a pure state has no negative values if and only if the state is Gaussian (in the finite case this statement holds only for odd dimensions $d$ \cite{gross2006hudson}). For example, the Weyl function of a coherent state is $\langle w|D(z)|w\rangle
   =e^{-\frac{1}{2}| z| ^2}e^{2 i
   \text{Im} z w^*}$, i.e. a Gaussian centered around $z=0$ that includes a phase, while the Wigner function is positive: it is a real Gaussian centered around $w$.

A straightforward calculation, repeating the steps taken for the qudit system, and using the SWAP relation  \eqref{eq:multi qmode SWAP}, immediately shows that
\begin{eqnarray}
   \int {\Pi _id^2z_i}p_{\psi}\left(z_1,\text{..},z_N\right)=|
   \langle \psi |\psi \rangle | ^2=1
\end{eqnarray}
therefore  $p_{\psi}(z_1,\text{..},z_N)$, the absolute value square of the Weyl function squared is a probability distribution on the set of displacements.

Starting from this point, there are many ways of defining a quantification of magic for CV states. As usual, due to the continuous nature of the distribution, there is a subtlety in defining appropriate entropy. One natural possibility is to use the Shannon differential entropy 
\begin{eqnarray}
  S_{Weyl}(\ket{\psi}) =  -\int {\Pi _id^2z_i}p_{\psi}log p_{\psi}
\end{eqnarray}
Crucially, we have that $S_{Weyl}(\ket{\psi})=S_{Weyl}(U \ket{\psi})$ for Gaussian operations $U$.

%
%

\begin{acknowledgments}
The work of IK was supported in part by the NSF grant DMR-1918207. I would like to thank A. Hamma, O. Pfister, M. Wampler and E. Altman for useful discussions.
\end{acknowledgments}

\bibliography{CVMagic.bib}
\bibliographystyle{unsrt}

\end{document}